\documentclass[aps,prx,superscriptaddress,reprint,showpacs,amssymb,amsmath,floatfix,citeautoscript]{revtex4-1}

\usepackage{graphicx}
\usepackage{dcolumn}
\usepackage{bm}
\usepackage{xcolor}
\usepackage[export]{adjustbox}
\usepackage{float}
\usepackage{tabularx}
\usepackage{multirow}
\usepackage{amsmath}
\usepackage{braket}
\usepackage{array}
\usepackage{ulem}

\newcolumntype{C}[1]{>{\centering\let\newline\\\arraybackslash\hspace{0pt}}m{#1}}

\begin{document}

\preprint{APS/123-QED}

\title{Singlet magnetism in intermetallic UGa$_2$ unveiled by inelastic x-ray scattering}

\author{Andrea~Marino}
     \affiliation{Max Planck Institute for Chemical Physics of Solids, N{\"o}thnitzer Stra{\ss}e 40, 01187 Dresden, Germany}
\author{Martin~Sundermann}
     \affiliation{Max Planck Institute for Chemical Physics of Solids, N{\"o}thnitzer Stra{\ss}e 40, 01187 Dresden, Germany}
     \affiliation{PETRA III, Deutsches Elektronen-Synchrotron DESY, Notkestra{\ss}e 85, 22607 Hamburg, Germany}
\author{Denise~S.~Christovam}
     \affiliation{Max Planck Institute for Chemical Physics of Solids, N{\"o}thnitzer Stra{\ss}e 40, 01187 Dresden, Germany}
\author{Andrea Amorese}
     \altaffiliation{ASML Netherlands B.V., De Run 6501, 5504 DR, Veldhoven, The Netherlands}
	 \affiliation{Max Planck Institute for Chemical Physics of Solids, N{\"o}thnitzer Stra{\ss}e 40, 01187 Dresden, Germany}
     \affiliation{Institute of Physics II, University of Cologne, Z\"{u}lpicher Str. 77, 50937 Cologne, Germany}
\author{Chun-Fu~Chang}
     \affiliation{Max Planck Institute for Chemical Physics of Solids, N{\"o}thnitzer Stra{\ss}e 40, 01187 Dresden, Germany}
\author{Paulius~Dolmantas}
     \affiliation{Max Planck Institute for Chemical Physics of Solids, N{\"o}thnitzer Stra{\ss}e 40, 01187 Dresden, Germany}
\author{Ayman~H.~Said}
     \affiliation{Advanced Photon Source, Argonne National Laboratory, 9700 S Cass Ave, Lemont, IL 60439, USA}
\author{Hlynur~Gretarsson}
     \affiliation{PETRA III, Deutsches Elektronen-Synchrotron DESY, Notkestra{\ss}e 85, 22607 Hamburg, Germany}
\author{Bernhard~Keimer}
     \affiliation{Max Planck Institute for Solid State Research, Heisenbergstra{\ss}e 1, D-70569 Stuttgart, Germany}
\author{Maurits~W.~Haverkort}
     \affiliation{Institute for Theoretical Physics, Heidelberg University, Philosophenweg 19, 69120 Heidelberg, Germany}
\author{Alexander~V.~Andreev}
     \affiliation{Institute of Physics, Academy of Sciences of the Czech Republic, Na Slovance 1999/2, 182 21 Prague 8, Czech Republic}
\author{Ladislav~Havela}
     \affiliation{Department of Condensed Matter Physics, Faculty of Mathematics and Physics, Charles University, Ke Karlovu 5, 121 16 Prague 2, Czech Republic}
\author{Peter~Thalmeier}
     \affiliation{Max Planck Institute for Chemical Physics of Solids, N{\"o}thnitzer Stra{\ss}e 40, 01187 Dresden, Germany}
\author{Liu~Hao~Tjeng}
     \affiliation{Max Planck Institute for Chemical Physics of Solids, N{\"o}thnitzer Stra{\ss}e 40, 01187 Dresden, Germany}
\author{Andrea~Severing}
     \affiliation{Max Planck Institute for Chemical Physics of Solids, N{\"o}thnitzer Stra{\ss}e 40, 01187 Dresden, Germany}
      \affiliation{Institute of Physics II, University of Cologne, Z\"{u}lpicher Str. 77, 50937 Cologne, Germany}
\date{\today}

\begin{abstract}
Using high resolution tender-x-ray resonant inelastic scattering and hard-x-ray non-resonant inelastic scattering beyond the dipole limit we were able to detect electronic excitations in intermetallic UGa$_2$ that are highly atomic in nature. Analysis of the spectral lineshape reveals that the local $5f^2$ configuration characterizes the correlated nature of this ferromagnet. The orientation and directional dependence of the spectra indicate that the ground state is made of the $\Gamma_1$ singlet and/or $\Gamma_6$ doublet symmetry. With the ordered moment in the $ab$ plane, we infer that the magnetism originates from the higher lying $\Gamma_6$ doublet being mixed with the $\Gamma_1$ singlet due to inter-site exchange, qualifying UGa$_2$ to be a true quantum magnet. The ability to observe atomic excitations is crucial to resolve the on-going debate about the degree of localization versus itineracy in U intermetallics.    
\end{abstract}

\maketitle

\section{Introduction}
Actinide intermetallics show a wealth of fascinating phenomena that includes heavy-fermion behavior, hidden order or unconventional magnetism, unconventional superconductivity, the combination of ferromagnetism and superconductivity\,\cite{Pfleiderer2009,Oppeneer2010,Mydosh2011}, orbital multicomponent\,\cite{Joynt2002}, or spin-triplet superconductivity\,\cite{Ran2019a, Ran2019b, Jiao2020, Hayes2021} with unusual topological properties\,\cite{Sato2017}. It is generally understood that those complex emergent properties originate from the intricate interplay of band formation and correlations involving the $5f$ electrons. It is, however, far from clear how to describe quantitatively the electronic structure of these systems, for example, whether an itinerant approach \cite{Oppeneer2010,Mydosh2011} or an embedded impurity model which includes explicitly the local degrees of freedom \cite{Haule2009,Kung2015,Miao2020} would be the better starting point. The main problem is that many intermetallic uranium compounds, perhaps with the exception of UPd$_3$\,\cite{osborn1990, le2011}, do not exhibit excitations in their inelastic neutron scattering data. It is therefore challenging to understand if remnants of atomic-like states are at all present in these compounds, let alone to pinpoint which multiplet and/or crystal-field  state is actually occupied.

\begin{figure*}[t!]
	\center
	\includegraphics[width=0.85\linewidth]{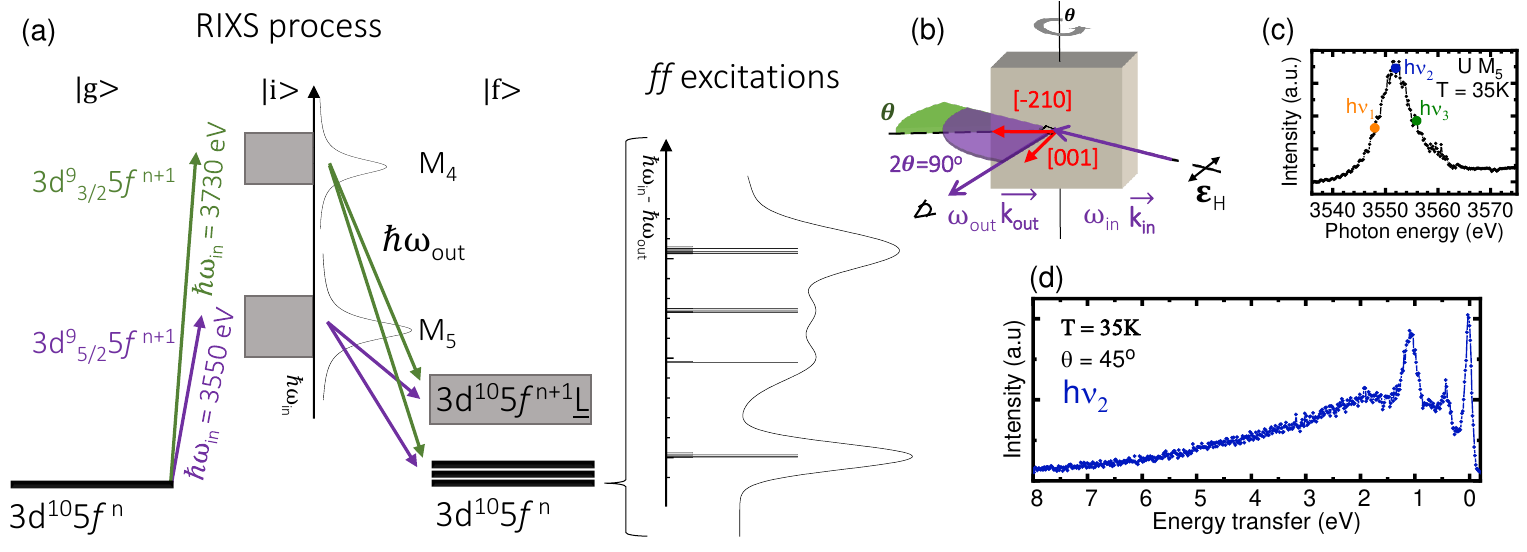}
\vspace{-0.3cm}
	\caption{(a) RIXS process at the U $M_{4,5}$ edge for a 5$f^n$ ground state. (b) Scattering geometry of the RIXS experiment. The scattering angle is kept at $2\Theta=90^o$. (c) Fluorescence-yield $M_5$ XAS spectrum of UGa$_2$. Large dots mark the photon energies where RIXS is measured ($h\nu_1$, $h\nu_2$, and $h\nu_3$). (d) $M_5$ RIXS spectrum of UGa$_2$ acquired at $h\nu_2$ and $\theta=45^\circ$.}
	\label{fig:Fig1}
\end{figure*}

Here we investigate UGa$_2$ as a representative case for many metallic U compounds in which the relative importance between itinerancy and localization is at issue in explaining the physical properties. UGa$_2$ crystallizes in the hexagonal AlB$_2$ structure (space group  P6/\textit{mmm})\,\cite{makarov1956}, with the U-U distances well above the Hill limit\,\cite{hill1970}, and orders ferromagnetically below $T_c$\,=\,125\,K with a small orthorhombic distortion\,\cite{andreev1979}. The moments are aligned in the $ab$-plane along the $a$ crystallographic direction.  The size of the uranium moment as determined by neutron diffraction\,\cite{lawson1985} and magnetization\, \cite{andreev1978, honma2000, kolomiets2015} measurements amounts to about $3\,\mu_B$, quite a high value as compared to other magnetically ordered uranium intermetallics, and suggests a more localized nature of the 5\textit{f} states in this binary. Inelastic neutron scattering, however, did not find crystal-field excitations; only magnons below 10\,meV were observed\,\cite{kuroiwa1993}. Attempts to explain the magnetism have been based on local $f^2$ and $f^3$ charge configurations \cite{radwanski1995,divis1996,richter1997,honma2000}, and on approaches that include itinerancy \cite{divis1996,chatterjee2021}. De Haas\,-\,van Alphen \cite{honma2000} and photoemission \cite{fujimori2019,kolomiets2021} experiments indicate that UGa$_2$ is neither localized nor itinerant. Spectroscopically, photoemission experiments are also not conclusive: core level data on bulk samples were interpreted as indicative for the localized nature of the 5\textit{f} states, based on the satellite structure of the U\,4\textit{f} core level that looks very different from that of itinerant UB$_2$\,\cite{schneider1981,fujimori2019}. On the other hand, data on UGa\textsubscript{2} films\,\cite{gouder2001} seem to support itinerancy, based on the fact that the satellites appear at different energy positions than in prototypical UPd$_3$.  

The observation of multiplets would provide direct evidence of the presence of atomic-like states. Furthermore, multiplets are a unique fingerprint of the configuration that determines the symmetry. Here resonant inelastic x-ray scattering (RIXS) is the ideal method because it covers a wide range of energy transfers. Already in 2006 Kvashnina \textit{et al.} and Burotin \textit{et al.} used RIXS at the U\,$O$-edge ($\approx100$\,eV) to distinguish valence states in semiconducting UO$_2$ and other U and actinide oxides\,\cite{kvashnina2006,butorin2011,butorin2013}, and Wray \textit{et al.} and Liu \textit{et al.} reported excitations in the $O$-edge RIXS spectra of intermatallic U compounds\,\cite{Wray2015,Liu2022}. However, the signal to background ratio of these $f$-$f$ excitations is very small at the U $O$-edge because of the strong elastic tail in the extreme ultraviolet. A recent publication of soft RIXS data at the U\,$N_4$ edge ($\approx778$\,eV), also of UO$_2$\,\cite{lander2021}, encouraged some of the authors of this manuscript to repeat the $N_4$-edge experiment with the same experimental set-up for the intermetallic large moment antiferromagnet UNi$_2$Si$_2$. The result was discouraging, with absolutely no inelastic intensity observed \cite{privatecomm2}. Another trial at the N$_5$-edge ($\approx736$\,eV) of the hidden order compound URu$_2$Si$_2$ gave the same negative result \cite{privatecomm1}. Kvashnina \textit{et al}., on the other hand, reported tender x-ray RIXS experiments at the U\,$M_4$ ($\approx3730$\,eV) and $M_5$ ($\approx3552$\,eV) edge with a resolution of 1\,eV for UO$_2$, and for the two intermetallic compounds UPd$_3$ and URu$_2$Si$_2$. Distinct excitations are observed at about 3\,-\,7\,eV (valence band into unoccupied 5$f$ states) and 18-20\,eV (U\,5$f$ to U\,6$p$), both at the $M_4$ and $M_5$ edge \cite{kvashnina2017}. These data show that the realization of \textit{high-resolution tender} RIXS at the U $M$-edges is the most promising direction to aim at, not only because of the expected stronger signal, but also because the tender x-ray regime does not require cleaving; it would even allow the confinement of samples. The latter would be a great advantage when performing experiments on U and especially actinide containing samples.


Here we utilize this new spectroscopic tool, namely \textit{high-resolution tender} RIXS at the U\,$M_5$ edge to tackle the origin of the magnetism in UGa$_2$.  With tender RIXS, we were able to detect  pronounced atomic multiplet states in the intra-valence band excitation spectrum of UGa$_2$. We also present hard x-ray core-level non-resonant inelastic scattering data (NIXS, also known as x-ray Raman) in the beyond-dipole limit at the U\,$O_{4,5}$-edge, confirming the RIXS result. Also in the high energy NIXS spectrum we observed states that are highly atomic in nature. Our analysis ultimately indicates that UGa$_2$ is a singlet ferromagnet.

\section{Methods}
\subsection{High resolution tender RIXS at U\,$M_5$-edge}

In a U\,$M_{4,5}$-edge RIXS experiment, see Fig.\,\ref{fig:Fig1}\,(a), a $3d$ core electron is excited from the 3$d^{10}$5$f^n$ ground state $|g \rangle$ into the 5$f$ shell by the absorption of incoming photons at the $M_5$ (3552\,eV) or $M_4$ (3730\,eV) edge, leading to 3$d^{9}_{5/2}$5$f^{n+1}$ or 3$d^{9}_{3/2}$5$f^{n+1}$ intermediate states $|i \rangle$, respectively. The subsequent de-excitation of the $3d$ core can be into the ground state (elastic peak), into an excited state of the same local charge configuration ($ff$ excitations, phonons, magnons)\,\cite{amorese2016, brookes2018, amorese2018, amorese2019}, or into an excited state of a different charge configuration (charge transfer excitations)\,\cite{nakazawa1996, butorin1996, dallera2001}. 

Fig.\,\ref{fig:Fig1}\,(b) depicts the experimental geometry where the scattering angle is set at 2$\Theta=90^\circ$ to minimize the elastic intensity. The UGa$_2$ samples used for the experiments were grown with the Czochralski method \cite{kolomiets2015} and their surface is the $ab$ plane, i.e. it has the [001] orientation.

High resolution tender RIXS was performed at the Max-Planck RIXS end station (IRIXS) of the P01 beamline at Petra III/DESY in Hamburg. The instrument is unique, since it allows to perform RIXS experiments with tender x-rays (2.5\,-\,3.5\,keV) and good resolution\,\cite{gretarsson2020}. For example, a resolution of 100\,meV can be achieved at the Ru $L_3$-edge at 2840\,eV. The IRIXS beamline uses the hard x-ray set-up\,\cite{gretarsson2020}. For the U\,$M_5$ edge at 3550\,eV a diced quartz waver (112), pressed and glued on a concave Si lens has been used as analyzer crystal\,\cite{said2018}. The instrument 150\,meV Gaussian response function at the U\,$M_5$ edge is estimated by measuring a carbon tape. The experiment was performed with horizontal polarization of the the incident photons, a scattering angle 2$\Theta=90^o$ to minimize elastic intensity and sample angles of $\theta=20^\circ, 45^\circ$ and $80^\circ$ (see Fig.\,\ref{fig:Fig1}(b)). Temperature was kept at 35\,K.

\subsection{NIXS with hard x-rays at U\,$O_{4,5}$-edge}
NIXS with hard x-rays (10\,keV) and large momentum transfer is dominated by higher-than-dipole transitions\,\cite{schuelke2007, haverkort2007}, which are more excitonic in contrast to the dipole contribution\,\cite{gordon2007, bradley2010, caciuffo2010, gupta2011}. The direction of the momentum transfer \textbf{$\vec{q}$} in NIXS plays an analogous role as the electric field vector \textbf{$\vec{E}$} in XAS and is sensitive to the symmetry of the crystal-field ground state.

The experiments are performed at the Max-Planck NIXS  end stations of the P01 beamline at Petra III/DESY in Hamburg. A sketch and description of the NIXS experimental setup is shown in Fig. 2 of \cite{sundermann2017}. 10\,keV photons are used. The average scattering angle $2\Theta$ at which the analyzers are positioned is $\approx$\,150\textsuperscript{o}, thus yielding a momentum transfer of $|$\textbf{$\vec{q}$}$|$=(9.6$\pm$0.1)\,\AA\textsuperscript{-1} at elastic scattering. An instrumental energy resolution of about 0.8\,eV FWHM is achieved. The sample is kept in a vacuum cryostat at $T=5$\,K. The O\textsubscript{4,5} edge of U is measured with momentum transfer \textbf{$\vec{q}$} parallel to the \textit{a} and \textit{c} crystallographic directions. 

\section{Results}
Fig.\,\ref{fig:Fig1}\,(c) shows the experimental U\,$M_5$-edge x-ray absorption (XAS) spectrum of UGa$_2$. The large dots mark the photon energies used in this RIXS study,  $E_\text{res}$-4\,eV ($h\nu_1$), $E_\text{res}$ ($h\nu_2$), and $E_\text{res}$+4\,eV ($h\nu_3$) with $E_\text{res}$ = 3552\,eV. In Fig.\,\ref{fig:Fig1}\,(d) the RIXS spectrum of UGa$_2$ is displayed for a wide energy range up to 8\,eV energy transfer taken at the $M_5$ resonance ($h\nu_2$) with the sample angle of $\theta=45^\circ$. The spectrum exhibits sharp peaks below 2\,eV that are on top of a broad feature that arises most likely from charge transfer excitations. The sharp peaks are very typical of local atomic-like excitations.

\begin{figure}[t!]
	\center
	\includegraphics[width=0.85\linewidth]{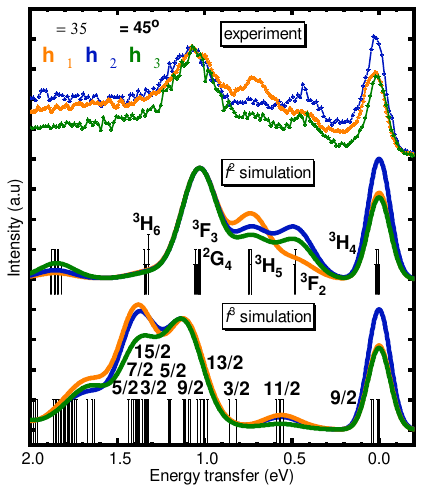}
	\caption{$M$\textsubscript{5} RIXS spectra of UGa\textsubscript{2} acquired at $h\nu_1$, $h\nu_2$, and $h\nu_3$ with a sample angle $\theta=45^\circ$ as compared to simulations using an $5f^2$ and $5f^3$ configuration.}
	\label{fig:Fig2}
\end{figure}

Fig.\,\ref{fig:Fig2} shows a close-up of the first 2\,eV of RIXS spectra that were measured with different incident energies, $h\nu_1$, $h\nu_2$ and $h\nu_3$. The data are normalized to the peak at 1.05\,eV. The intensities of the peaks vary considerably with the incoming photon energy so that three inelastic excitations at 0.44, 0.70 and 1.05\,eV can be identified. We assign these to intermultiplet $ff$ transitions since the energies are too high for magnons and crystal-field excitations.

Full atomic multiplet calculations assuming a $5f^2$ and alternatively a $5f^3$ configuration were carried out to simulate the spectra. For this, the \textit{Quanty} code \cite{haverkort2016} was used with the atomic values of the the spin-orbit constant and the $5f$-$5f$ and $3d$-$5f$ Slater integrals from the Atomic Structure Code by Robert D. Cowan \cite{cowan1981} as input parameters, whereby the spin orbit constant was reduced by 10\% and the Slater integrals by 45\% in order to take configuration interaction effects and covalence into account\,\cite{Tanaka1994,Groot2008,Agrestini2017}. These are typical reduction factors for uranium compounds\,\cite{lander2021}. A crystal-field (CF) potential was always considered. CF parameters were taken from fits to the magnetic susceptibility or magnetization, for the $5f^2$ from Ref.\,\cite{richter1997} and for the $5f^3$ configuration from Ref.\,\,\cite{honma2000,radwanski1995}, or constructed to test different CF ground state wave functions. Furthermore, a Lorentzian broadening of about 6\,eV in the intermediate state was used, based on the width of the $M_5$ XAS spectrum, and a Gaussian broadening of 150\,meV to account for the experimental resolution.

Fig.\,\ref{fig:Fig2} also shows the simulation for the $5f^2$ and for the $5f^3$ configuration. The calculation based on the $5f^2$ reproduces the experimental data in terms of peak positions as well as variation of the peak intensities with incident energy. The vertical lines represent the histogram of the multiplet states and provide a straightforward labeling of the peaks. The $5f^3$ simulation, on the other hand, does not reproduce the experimental data. It turns out that no matter how the reduction factors are tuned, an agreement cannot be achieved for 5$f^3$ (see Appendix\,\ref{suppl:RIXSf3}). Hence we conclude, the atomic-like states in UGa$_2$ are given by the $5f^2$ configuration. 

Next we determine the CF symmetry of the ground state. Here we ignore the slight orthorhombic distortion below $T_c$ \cite{andreev1979} since it is only a very small magnetostriction correction to the hexagonal crystal-field analysis. In \textit{D\textsubscript{6h}} the hexagonal CF splits the nine-fold degenerate \textit{J}\,=\,4 Hund's rule ground state of the U $5f^2$ configuration into three singlets and three doublets. These be written in the $J_z$ representation as:

\begin{align}
    \Gamma_1 & =| 0\rangle \\
    \Gamma_3 & =\frac{1}{\sqrt{2}}|+3\rangle + \frac{1}{\sqrt{2}}|-3\rangle \\
    \Gamma_4 & =\frac{1}{\sqrt{2}}|+3\rangle - \frac{1}{\sqrt{2}}|-3\rangle \\ \Gamma_5^1 & =\sin{\phi}|\pm4\rangle + \cos{\phi}|\mp2\rangle \\
    \Gamma_5^2 & =\cos{\phi}|\pm4\rangle - \sin{\phi}|\mp2\rangle \\
    \Gamma_6 & =|\pm1\rangle
\end{align}

 \begin{figure}[t!]
	\center
	\includegraphics[width=1\linewidth]{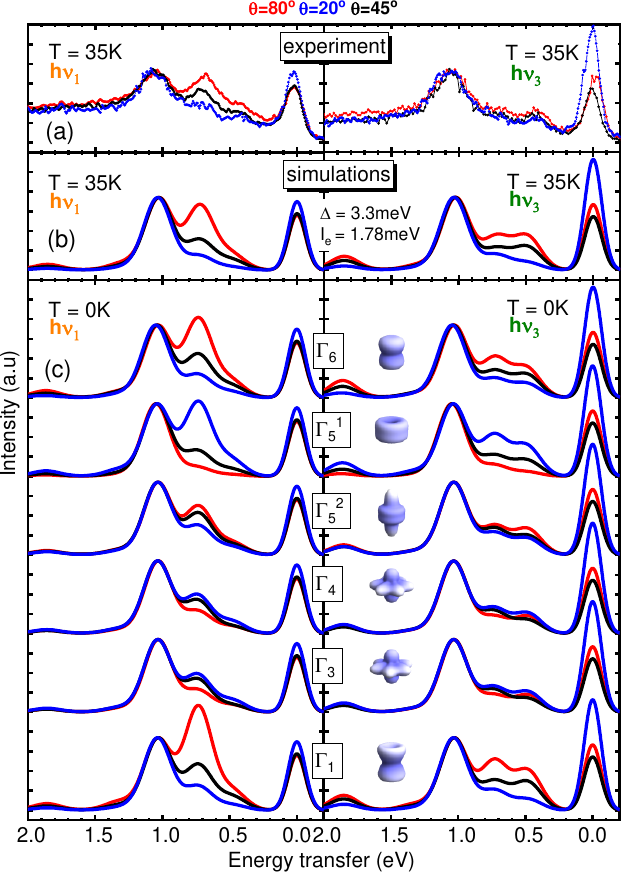}
\vspace{-0.2cm}
	\caption{(a)   Experimental RIXS spectrum of UGa$_2$ acquired at $h\nu_1$ and $h\nu_3$ with 
		sample angles $\theta=80^\circ$,\,$45^\circ$\,and\,$20^\circ$. (b) Simulated 35\,K RIXS 
		spectra for an $5f^2$ configuration with the crystal-field parameters of Richter 
		$et$\,$al.$ \cite{richter1997} and a molecular field $I_e$ of 1.78\,meV. 
		(c) Simulated 0\,K RIXS spectra for each of the six crystal-field states of the $5f^2$.}
	\label{fig:Fig3}
\end{figure}

Although the CF splitting is below the resolution limit of the present RIXS experiment, it is possible to obtain information about the ground state symmetry by measuring the orientation dependence of the scattering signal\,\cite{Kotani2001}. The two panels of Fig.\,\ref{fig:Fig3}(a) show the RIXS spectra for two incident energies ($h\nu_1$ and  $h\nu_3$), whereupon each energy was measured for the three sample angles $\theta=80^\circ$,\,$45^\circ$\,and\,$20^\circ$. The $\theta$ rotation is in the [001]-[-210]  plane, see Fig.\,\ref{fig:Fig1}(b). The intensities are again normalized to the peak at 1.05\,eV energy transfer. A pronounced orientation dependence can be seen in the spectra for $h\nu_1$, that has almost disappeared for $h\nu_3$.

Again the data are compared to the full multiplet calculations. In Fig.\,\ref{fig:Fig3}(b) we start with the calculations using the $5f^2$ crystal field parameters of Richter $et$\,$al.$ \cite{richter1997}. With this set of parameters, the ground state is the $\Gamma_1$, the first excited state the $\Gamma_6$ at 3.3\,meV, the second excited state the $\Gamma_5^1$ with $\sin{\phi}$\,=\,0.81 at 5.9\,meV, and all other states at 13 meV or higher. The calculations were performed for T\,=\,35\,K and include a molecular field $I_e$ of 1.78\,meV as will be explained later. We observe that the calculations reproduce the experiment well: the simulation captures the strong orientation dependence for $h\nu_1$ in the correct sequence and its decrease for $h\nu_3$.

To understand whether a different order of CF states would also be able to reproduce the experiment, we calculate the spectra for different CF ground states. To this end, we tuned slightly the CF parameters such that the desired CF state becomes the ground state and then we carried out the calculation for T\,=\,0\,K. The results are displayed in Fig.\,\ref{fig:Fig3}(c). We observe that a $\Gamma_1$ ground state, or a $\Gamma_6$, or also a $\Gamma_5^2$ with $\cos{\phi}$\,$\approx$\,1 have the correct trend in the orientation dependence for $h\nu_1$ and its reduction at $h\nu_3$. A $\Gamma_3$, $\Gamma_4$, or $\Gamma_5^{1}$ as ground state, on the other hand, produces an orientation dependence that is opposite to the experiment so that these three states can be excluded. The NIXS experiment below will show that also the $\Gamma_5^{2}$ cannot be the ground state. We thus conclude that the good simulation is based on the strong orientation dependence provided by the $\Gamma_1$ or the $\Gamma_6$  low lying states, which gets counteracted at 35\,K by the Boltzmann occupation of a higher lying state with opposite orientation dependence, such as the $\Gamma_5^1$.
\begin{figure}[t!]
	\center
	\includegraphics[width=1\linewidth]{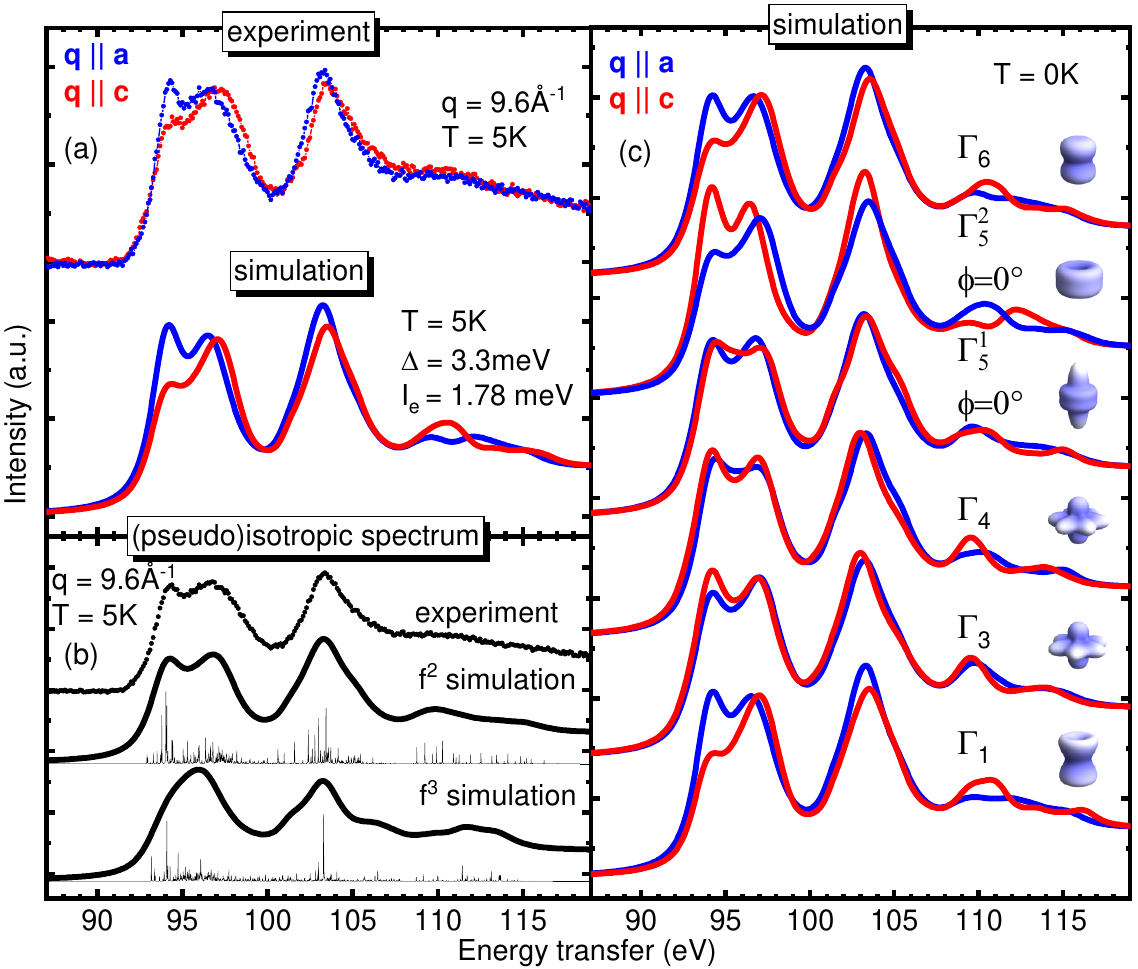}
\vspace{-0.1cm}
	\caption{(a) NIXS spectra for \textbf{$\vec{q}$}\,$||$\,$a$ and \textbf{$\vec{q}$}$||$\,$c$
		together with the simulation using the crystal field parameters of Richter 
		$et$\,$al.$ \cite{richter1997} and a molecular field $I_e$ of 1.78\,meV. 
		(b) (Pseudo)isotropic data and full multiplet simulations without crystal-field 
		for the U\,$f^2$ and U\,$f^3$ configurations. (d) Simulated NIXS spectra for each of the 
		six crystal-field states of the $f^2$ together with the corresponding charge densities.}
	\label{fig:Fig4}
\end{figure} 

Fig.\,\ref{fig:Fig4}(a) shows the $O_{4,5}$ edge NIXS data of UGa\textsubscript{2} at 5\,K for \textbf{$\vec{q}$}\,$||$\,$c$ and \textbf{$\vec{q}$}\,$||$\,$a$, revealing a strong directional dependence. In Fig.\,\ref{fig:Fig4}(b) the (pseudo) isotropic U $O_{4,5}$ NIXS spectrum, constructed from $I_{iso}=(I_{\textbf{q}||c}+2I_{\textbf{q}||a})/3$, is displayed and compared to atomic simulations without considering the crystal-field Hamiltonian. The Slater integrals for the $5f$-$5f$ and $5d$-$5f$ Coulomb interactions are reduced by about 40\% with respect to their atomic values. The value of the momentum transfer in the simulation is set to $|$\textbf{$\vec{q}$}$|\,=\,11.1$\,\AA\textsuperscript{-1} in order to account for the U $5f$ radial wave function in the solid being different from the calculated atomic value. An arctangent type of background is added to account for the edge jump. A Gaussian broadening of 0.8\,eV and a Lorentzian broadening of 1.3\,eV account for instrumental resolution and lifetime effects, respectively. The simulations are performed both for an $5f^2$ and $5f^3$ configuration and also here only the $f^2$ simulation reproduces the experimental lineshape, whereas the $f^3$ does not (see Appendix\,\ref{suppl:f3_NIXS}). This finding is fully consistent with the RIXS results.

Focussing now on the directional dependence, we calculate the spectra for each of the six possible crystal-field states. The results are displayed in Fig.\,\ref{fig:Fig4}(c). Comparison with experiment immediately excludes the $\Gamma_3$ and $\Gamma_4$ singlets, as well as the $\Gamma_5$ doublets for any range of the parameter\;$\phi$ (see equations above). For the $\Gamma_5^{(1,2)}$ doublets only the extreme cases of $\phi=0^o$ are shown, since the spectra for all other $\phi$ values fall between these two extremes. The $\Gamma_1$ singlet and the $\Gamma_6$ doublet, on the other hand, show the same directional dependence as the experiment, thus confirming the RIXS results.

\section{Discussion}
The above NIXS and RIXS results find that the 5$f^2$ configuration dominates the local electronic structure of UGa$_2$ and that the symmetry of the CF ground state is either given by the $\Gamma_1$ singlet and/or $\Gamma_6$ doublet. However, we can further exclude the $\Gamma_6$ doublet as ground 
state because it would yield an ordered moment along $c$ and not in the $ab$-plane. Hence, the $\Gamma_1$ singlet state must be the lowest one in energy. Yet, the $\Gamma_6$ is also a necessary ingredient for the magnetism in UGa$_2$ as we will discuss in the following.

In a conventional local moment magnet the non-vanishing temperature independent moments are present at each lattice site and then order spontaneously at the transition temperature creating a self-consistent molecular field. This is basically a classical concept modified only by the influence of semiclassical quantum fluctuations which reduce the size of the ordered moment by a modest amount. A $\Gamma_1$ ground state would not carry a local moment so that the semiclassical picture of magnetic order does not apply, it rather must be classified as a true quantum magnet where the creation of the local moments and their ordering appears spontaneously at $T_c$. This mechanism of induced magnetic order is caused by the non-diagonal mixing of $\Gamma_1$ with excited $\Gamma_6$ states due to the effective inter-site exchange coupling that forms the true ground state superposition below the ordering temperature. Induced quantum magnetism in singlet ground state systems has been explored in $d^4$ transition metal\,\cite{giniyat2013,jain2017} or 4$f^2$ Pr materials (see Ref.\,\cite{Thalmeier2021} and references therein). In these cases the presence of multiplets is clear. Singlet magnetism is however rarely recognized in U compounds\,\cite{Miao2019,Thalmeier2021,Thalmeier2002,Thalmeier2005,amorese2020}, where pinpointing the U\,5$f^2$ configuration is already challenging (see also Appendix\,\ref{suppl:inducedmoment}).

Looking at the simple structure of CF states, we realize that indeed $\Gamma_6$ is the only possible excited state that has non-vanishing mixing matrix elements $\langle\Gamma_1|J_{x}|\Gamma_6\rangle$ for the in-plane total angular momentum operators (not for $J_z$) so that there can be no coupling to any other state when we restrict to the Hilbert space of the ground state multiplet ($J$\,=\,4). This explains naturally that the ordered moment must lie in the hexagonal plane and at the same time the anisotropy of the paramagnetic susceptibility. For the induced moment mechanism of magnetic order to work, i.e. to produce a finite ordering temperature, the effective exchange $I_e$ must surpass a critical value. Here $I_e$ is the Fourier transform $I(\textbf{q})$ of the inter-site coupling $I_{ij}$ at the ordering vector $\textbf{q}$ where $I(\textbf{q})$ is at its maximum. In a singlet ground state system with a $\Gamma_1$\,-\,$\Gamma_6$ splitting energy $\Delta$ a spontaneous induced moment can only appear when the control parameter $\xi$ is larger than 1, with $\xi$\,=\,$2 \alpha^2$\,$I_e$/$\Delta$\,\cite{Thalmeier2002, Thalmeier2021} and $\alpha^2$\,=\,$\sum _\sigma$\,$|\langle\Gamma_1|J_{x}|\Gamma_{6,\sigma}\rangle|^2$, where $\sigma$ is the degeneracy index of the $\Gamma_6$ states, with numerical value $\alpha$\,=\,3.1. The saturation moment at zero temperature then is given by $m_0/(g_J\mu_B)$\,=\,$\langle J_{x} \rangle _0$\,=\,$\alpha \xi^{-1}(\xi^2-1)^{1/2}$ ($g_J = 0.8$) which vanishes when approaching the critical value from above $\xi \rightarrow 1^+$, and becomes equal to $\langle J_{x} \rangle _0 = \alpha$, that of a quasi-degenerate $\Gamma_1$\,-\,$\Gamma_6$ system, when $\xi \gg 1$, i.e. where the effective exchange strongly dominates over the splitting. In UGa$_2$ the moment of 3\,$\mu_B$\,/\,U  is close to the latter case of the exchange dominated regime. See also Appendix\,\ref{suppl:inducedmoment}.
\begin{figure}[t!]
	\center
	\includegraphics[width=0.8\linewidth]{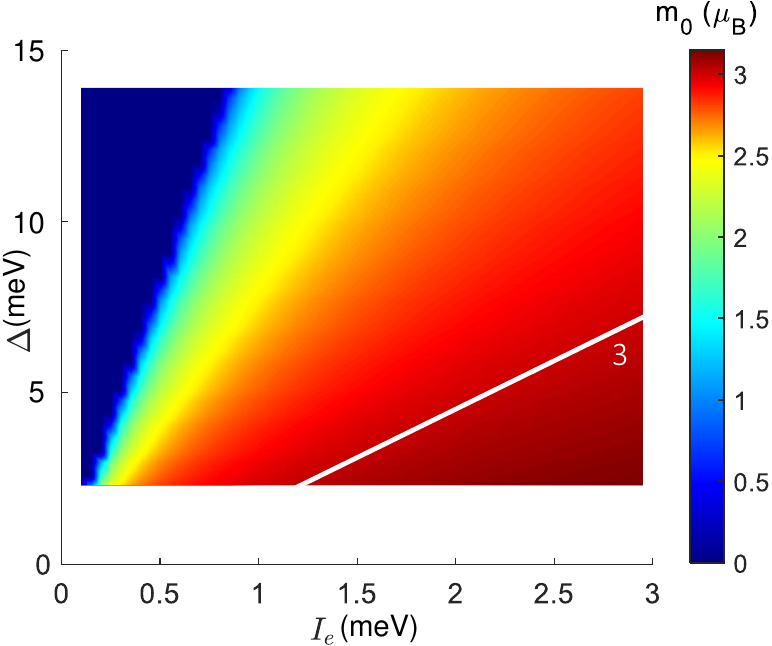}
\vspace{-0.3cm}
	\caption{Magnetic moment $m_0$ at zero temperature as a function of the $\Gamma_1-\Gamma_6$ splitting $\Delta$ and the effective exchange $I_e$. The white line denotes the contour with $m_0=3$\,$\mu_B$}
	\label{Fig4}
\end{figure}

The thermal occupation of higher levels has to be considered for the determination of the temperature dependence of $\langle J_{x} \rangle_T$ and $T_c$ as a function of the exchange $I_e$ and CF splitting $\Delta$. This can be done within our full multiplet calculation (including all angular momentum multiplets $J$ and their respective CF multiplets) by solving iteratively the selfconsistency equation $\langle J_{x,y} \rangle_T = \sum _n p_n \langle n | J_{x,y} | n \rangle$  where $E_n(\langle J_{x,y} \rangle _T)$ and $|n\rangle(\langle J_{x,y} \rangle _T)$ are the eigenenergies and eigenstates in the presence of the molecular field $I_e\langle J_{x,y} \rangle_T$ for the given values of multiplet model parameters. Here $p_n$\,=\,$Z^{-1}\exp(-E_n/T)$ with $Z=\sum_m \exp(-E_m/T)$ are the thermal level occupations. The saturation moment $m _0/\mu_B=g_J\langle J_{x} \rangle _0 $ may then be plotted as function of the splitting $\Delta$ and exchange $I_e$ as shown in Fig.\,\ref{Fig4}. Here the CF parameters from Ref.\,\cite{richter1997} (apart from the off-diagonal $A_6^6$) are scaled to modify the splitting $\Delta$. It would be interesting to measure $\Delta$ using Raman spectroscopy\,\cite{buhot2014}. For $\Delta$\,=\,3.3\,meV and $I_e$\,=\,1.78\,meV, corresponding to $m_0(\Delta,I_e)$\,$\approx$\,3\,$\mu_B$ (one point on the white line in Fig.\,\ref{Fig4}), we obtain $T_c \approx$\,125\,K, in agreement with experiment. For completeness, we finally examine the impact of the self consistent molecular field with $I_e$\,=\,1.78\,meV on the RIXS (see Fig.\,\ref{fig:Fig3}\,(b)) and NIXS (see Fig.\,\ref{fig:Fig4}\,(a)) spectra. Here the CEF scheme giving  $\Delta$\,=\,3.3\,meV and the Boltzmann population of excited states is also considered. We see that the molecular field has little impact on the spectra, since, although mixed, the $\Gamma_1$ and $\Gamma_6$ show very similar lineshapes by themselves.

\section{Conclusion}
In summary, with tender RIXS at the U $M_5$ edge and hard x-ray NIXS at the U $O_{4,5}$-edge we have unveiled the U\,$5f^2$ multiplets inUGa$_2$ and shown that the magnetism is determined by the U\,5$f^2$ configuration with a $\Gamma_1$ singlet ground state and a $\Gamma_6$ doublet nearby.  UGa$_2$, therefore, classifies as a 
quantum magnet. The origin of the induced magnetic order is due to the non-diagonal mixing of $\Gamma_1$ with excited $\Gamma_6$ states due to the effective inter-site exchange coupling below $T_c$.  

\section{Acknowledgements}
All authors thank C. Geibel and A.C. Lawson for fruitful discussions, and acknowledge DESY (Hamburg, Germany), a member of the Helmholtz Association HGF, for the provision of experimental facilities.  A.S. and A.A. benefited from support of the German Research Foundation (DFG), Project No. 387555779. L.H. and A.V.A. benefited from support of the Czech Science Foundation, Project No. 21-09766S. 

\section{Appendix}

\subsection{$f^3$ RIXS simulation with different values of the reduction factors}
\label{suppl:RIXSf3}
\begin{figure}[h!]
\center
\includegraphics[width=0.85\linewidth]{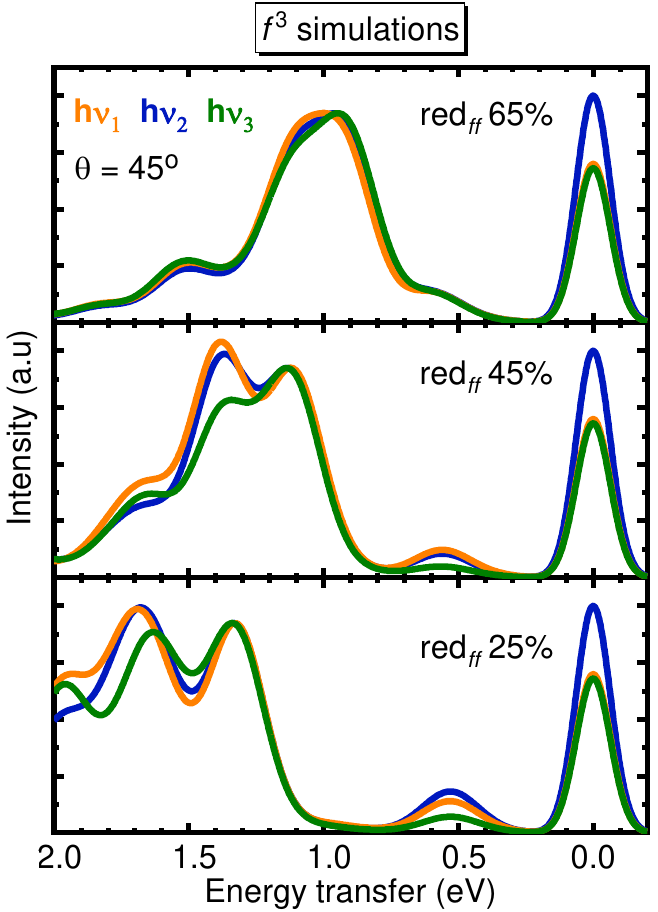}%
\caption{Ionic $f^3$ simulated RIXS for different values of the  $5f$-$5f$ reduction factors. }
\label{fig:Fig1_SI}
\end{figure}

\vfill\null

Fig.\,\ref{fig:Fig1_SI} shows simulated $f^3$ RIXS spectra with $5f-5f$ Slater integral reduced of 45\%, 40\% and 75\%. The spectra are calculated with incident energies $h\nu_1$, $h\nu_2$ and $h\nu_3$, and with the crystal-field parameters from \cite{radwanski1995}.

\subsection{$f^3$ isotropic NIXS simulation with different values of the reduction factors}

\label{suppl:f3_NIXS}

\begin{figure}[h!]
\center
\includegraphics[width=0.85\linewidth]{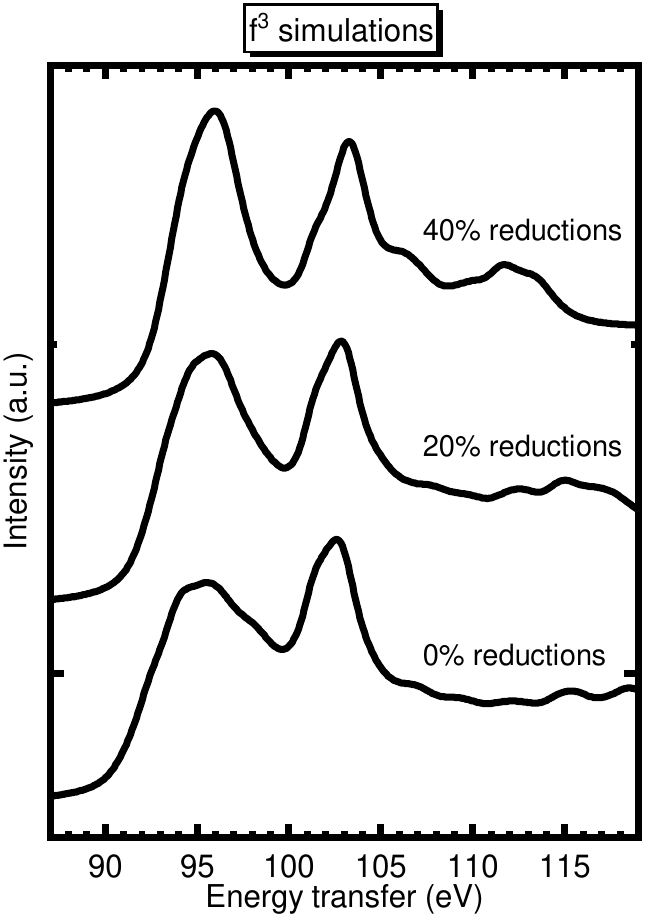}%
\caption{Ionic $f^3$ simulated isotropic NIXS for different values of the  $5f$-$5f$ and $5d$-$5f$ reduction factors. }
\label{fig:Fig2_SI}
\end{figure}

Fig.\,\ref{fig:Fig2_SI} shows simulated $f^3$ NIXS spectra with $5f$-$5f$ and $5d$-$5f$ Slater integral reduction to 60\%, 80\% and 100\%. The crystal-field is not included in the calculations.

\subsection{Crystal-field parameters and ground state symmetry}
\label{suppl:CFparameters}

 Table\,\ref{tab:table_CEF} summarizes the crystal-field parameters and the corresponding ground state symmetries used in the  RIXS calculations. For the $\Gamma_5$ states the relative $J_z=|\pm4\rangle$ and $J_z=|\mp2\rangle$ contribution is specified. The parameters giving a $\Gamma_1$ ground state are taken rom Ref.\,\cite{richter1997}.

\begin{table}[h]
    \centering
    \begin{tabularx}{\columnwidth}{|c|X|X|X|X|}
        \hline
        GS symmetry & \multicolumn{1}{c|}{$A_2^0$ (K)} & \multicolumn{1}{c|}{$A_4^0$ (K)} & \multicolumn{1}{c|}{$A_6^0$ (K)} & \multicolumn{1}{c|}{$A_6^6$ (K)} \\ \hline \hline
        $\Gamma_1$ & -220 & 680 & -100 & 1040 \\
        \hline
         $\Gamma_6$ & -1044.41 & 696.272 & -1856.72 & -812.317 \\
        \hline
        $\Gamma_3$ & -365.543 & -731.085 & 417.763 & -731.085 \\
        \hline
        $\Gamma_4$ & -292.434 & -841.328 & 300.557 & 731.085 \\
        \hline
        $\Gamma_5^1$ & \multirow{2}{*}{438.651} & \multirow{2}{*}{877.302} & \multirow{2}{*}{501.315} & \multirow{2}{*}{-877.302} \\
       {\tiny $0.991|\pm4\rangle + 0.128|\mp2\rangle$} & & & & \\
        \hline
        $\Gamma_5^2$ & \multirow{2}{*}{-626.644} & \multirow{2}{*}{-584.868} & \multirow{2}{*}{-313.322} & \multirow{2}{*}{522.204} \\
        {\tiny $0.075|\pm4\rangle - 0.997|\mp2\rangle$} & & & & \\
        \hline
    \end{tabularx}
    \label{tab:table_CEF}
\end{table}

\subsection{Basics of induced moment magnetism}
\label{suppl:inducedmoment}

In this work the magnetism of UGa$_2$ is interpreted in terms of a localized model consisting of $5f$ CEF states for $J=4$.  The on-site exchange  interaction (resulting from Anderson-type on-site hybridisation and Coulomb repulsion)
between conduction and f-electron is assumed to have been eliminated leading to an effective RKKY-interaction $I_{ij}$ between $5f$ states on different sites $i,j$. $I_e$ is the Fourier transform $I(\textbf{q})$ of the inter-site coupling $I_{ij}$ at the ordering vector $\textbf{q}$ where $I(\textbf{q})$ is at its maximum. Restricting to FM with $\textbf{q}=0$ and to nearest neighbor terms only, the effective interaction is given by $I_e=zI_{nn}$ (z=coordination number). If the 5f ground state were degenerate and carried an effective moment (i.e. having nonzero matrix elements of ${\bf J}$ within the multiplet) a quasiclassical ferromagnetic order would appear for any size of $I_e$ where moments are simply aligned at a temperature $T_C \sim I_e$. Here, however, the lowest 5f states are nonmagnetic singlet $\Gamma_1$ ground state and $\Gamma_6$ doublet excited state at energy $\Delta$. Due to their absent moments the FM order in UGa$_2$ can only appear through a more subtle mechanism called 'induced order'. This mechanism is well established for several 4f Pr and 5f U compounds with nonmagnetic low lying CEF states as in the present case. We refer to previous Refs.~\cite{grover:65,cooper:72,birgeneau:72,buyers:75,jensen:91,Thalmeier2002,Thalmeier2021} for the detailed discussion of the subject. Although the $\Gamma_1,\Gamma_6$ states do not carry a moment there are {\it nondiagonal} matrix elements $\alpha/\sqrt{2}=\langle \Gamma_1|J_x|\Gamma_{6\sigma}\rangle$ ($\sigma =1,2$) of in-plane dipolar moment $J_x$ (and similar for $J_y$) connecting
them accross the CEF gap $\Delta$. This means that n.n. inter-site interaction terms like $I_{ij}J_x(i)J_x(j)$ are able to
mix the excited state $\Gamma_6$ into the noninteracting ground state $\Gamma_1$ and form spontaneously a new
magnetic ground state at each site which is a superposition $|\Gamma'_1\rangle=u|\Gamma_1\rangle+v|\Gamma_6\rangle$ (and similar for the excited state). In this way the ground state moment appearance and its ordering happens simultaneously. The size of the ordered moment is then $\langle J_x\rangle = 2uv\alpha(n'_1-n_6')$ where $n'_{1,6}$ denote the thermal occupations of the CEF states which also depend on $\langle J_x\rangle$. This represents a molecular field equation for the induced moment $\langle J_x\rangle$. When temperature is lowered the occupation difference increases which may lead to a nonzero induced moment, provided the prefactor in the above equation is sufficiently large. This can be evaluated as a condition for
the control parameter $\xi=2\alpha^2I_e/\Delta >1$ to achieve a finite T$_C$ and a saturation moment at $T=0$ given by 
 $\langle J_x\rangle_0=\alpha\xi^{-1}(\xi^2-1)^{\frac{1}{2}}$. At zero temperature varying $\xi$ across the quantum critical point (QCP) $\xi=1$ we obtain a quantum phase transition from the paramagnetic $(\xi<1)$ to the (ferro-) magnetic $(\xi>1)$ state. In particular close to the QCP the induced moment quantum magnetism shows anomalous dependence of small saturation moment and low ordering temperature on the control parameter and is quite different from the quasiclassical magnetism
where the influence of quantum fluctuations on moment and transition temperature is moderate.


\providecommand{\noopsort}[1]{}\providecommand{\singleletter}[1]{#1}%

\end{document}